# Fault Diagnosis and Fault-Tolerant Operation Strategy for Triple-Port Hydrogen Production System

Shiqi Zhang, *Student Member, IEEE*, Yiyina Teng, Yuhang Sun, Fanqin Ding, Ning Wang, *Member, IEEE*, Xiaoqiang Guo, *Senior Member, IEEE*, Changchun Hua, *Fellow, IEEE*

***Abstract*—The multi-port converter offers significant advantages in sustainable energy-based hydrogen production, such as high energy density and wide application potential. As the core energy hub in these systems, the multi-port hydrogen converter demands high reliability. However, its numerous switches and complex operation increase the risk of failure. Currently, there are no effective fault diagnosis or fault tolerance solutions for these systems. Current fault diagnosis methods for two-port and multi-port converters rely on many sensors, have low accuracy, and can't handle multi-switch faults. Additionally, existing fault tolerance strategies fall short in maintaining stable power for the electrolyzer and don't allow capacity sharing among ports. This paper proposes a fault diagnosis method for open-circuit faults in a triple -port hydrogen converter and a corresponding fault-tolerant operating strategy. By optimizing a random forest algorithm with a sparrow search algorithm, high-precision multi-switch fault diagnosis is achieved using only three current sensors. Fault-tolerant operation of the system is achieved through an auxiliary switch after a failure. Additionally, power sharing between output ports can be realized. This ensures the efficient and stable operation of the hydrogen production system. The effectiveness of this solution has been validated through simulations and experiments.**

## I. INTRODUCTION

SUSTAINABLE energy serves as a vital foundation for human survival and social development, playing a crucial role in promoting the progress of modern economic society and improving the global climate due to the depletion of fossil fuels and environmental pollution [1, 2]. Hydrogen energy, as a novel sustainable resource, has become an indispensable part of future energy strategy due to its abundant reserves, high energy density, and clean efficiency [3, 4]. Hydrogen can be produced through electrolysis, thermal processes, and biochemical methods, among which water electrolysis is the most convenient [5, 6]. Water electrolysis for hydrogen production is considered a promising alternative for the future to reduce fossil fuel consumption and greenhouse gas emissions [7, 8].

The hydrogen production converter, a core component of the hydrogen production system, significantly influences the purity, efficiency, and lifespan of both the hydrogen produced and the electrolyzer [9]. The use of multi-port hydrogen production converter in hydrogen production technology has attracted considerable attention from experts both domestically and internationally and is considered a key area for future growth in the power industry [10]. Given the high-reliability requirements of electrolyzers [11, 12], fault diagnosis and fault-tolerant operation of multi-port hydrogen production converter are essential for ensuring the stability of sustainable energy-based hydrogen production [13], The flexibility of hydrogen as a controllable load [14], as well as for optimizing the power output, efficiency, and cost of the electrolyzer [15]. Compared to traditional hydrogen production converters, multi-port hydrogen converter feature more switches and complex operating modes, posing new challenges to the stable and efficient operation of hydrogen production systems [16].

In the hydrogen production converter, switching device faults primarily include short-circuit and open-circuit faults. While short-circuit faults are relatively straightforward to diagnose, open-circuit faults are more challenging to detect and can easily lead to system malfunction, posing significant safety risks [17]. Therefore, this paper focuses on the diagnosis of open-circuit faults. Existing research utilizes state estimators and normalized detection variables to identify faulty units and locate defective transistors based on residual voltage and current characteristics [18], however, the detection time is relatively long. Some studies also use the midpoint voltage of the neutral-point clamped bridge arm as a fault diagnosis signal, identifying faulty switches based on the average value and duty cycle of the midpoint voltage. However, this method requires two additional voltage sensors for midpoint voltage detection [19]. To improve noise immunity in diagnosis, a vector-based fault diagnosis method is proposed, which diagnoses faults by identifying the DC offset vector relationship of three-phase currents. However, it requires additional low-bandwidth current sensors on the transformer side [20]. Yang W. et al. propose a logic diagnosis method that uses the periodic average of the midpoint voltage of each bridge arm as a fault indicator, comparing the DC offset with a set threshold. However, this approach requires a relatively large number of sensors—specifically, three voltage sensors and two current sensors [17]. Some studies have improved diagnostic accuracy through a precise fault transistor identification algorithm based on a post-fault control reconfiguration (PFRC) and redundant leg (PFRR) strategy [21]. While PFRC can identify faulty transistors without the need for redundant legs, it has a longer identification time. Additionally, for a triple-active bridge (TAB), this method involves complex reconfiguration of the current path between the three ports, and the design of the redundant leg circuit is more challenging.

The stable operation of the hydrogen electrolyzer is crucial [22]. Fault tolerance in hydrogen production is essential for green hydrogen production. Implementing fault-tolerant control based on the specific fault location within the converter helps mitigate the impact of switch failures and restore the converter to a stable operating state. Generally, fault-tolerant strategies can be divided into fault isolation, control algorithm reconfiguration, and topology reconfiguration.

Existing research eliminates DC bias caused by faults through a method that adjusts the phase shift between ports based on fault modes or reconfigures the faulty bridge arm into a half-bridge [23], and a fault-tolerant strategy for a triple-port converter that reconfigures the switches on the same bridge arm as the faulty switch through a control algorithm to eliminate the port inductance current offset caused by the fault [17]. Some studies have added blocking capacitors on both sides of the transformer and used a blocking method to achieve fault tolerance by reconfiguring the path through the faulty phase switch [24]. Alternatively, a fault-tolerant control scheme based on a complementary switch blocking method is used, where the gate drive signal of the complementary switch is blocked when one switch experiences an open-circuit fault (OCF), achieving fault tolerance [19]. Other research describes using diodes in the faulty phase to block the complementary switch and reconfigure the modal current path for fault-tolerant operation [25].

After implementing the fault-tolerant control strategies mentioned above, the output power at the ports of the multi-port hydrogen production converter is significantly reduced, leading to a decrease in hydrogen production efficiency and an increase in costs for the electrolyzer. In severe cases, this can result in shutdowns and equipment damage. Furthermore, as shown in Table II, none of the existing fault-tolerant control strategies can achieve direct power sharing between the load ports of a triple-port hydrogen production converter.

This paper proposes a method for fault diagnosis and fault-tolerant control in multi-port hydrogen converter using auxiliary switches. The method leverages a sparrow search algorithm-optimized random forest algorithm for fault diagnosis, followed by the implementation of fault-tolerant operations based on the diagnostic results.

The main contributions of this paper are summarized as follows: 1) A fault diagnosis method based on a sparrow search algorithm-optimized random forest (SSA-RF) algorithm is proposed, enabling fault diagnosis with fewer sensors. 2) The method provides high diagnostic accuracy and speed and can diagnose multiple switch faults. 3) A fault-tolerant operation method using auxiliary switches is introduced, which is easy to implement and offers good responsiveness and robustness. 4) The proposed fault-tolerant method allows for power sharing between the load-side ports of a triple-port hydrogen production converter.

The rest of this paper is organized as follows: The second section analyzes the open-circuit fault characteristics of the triple-port hydrogen production converter and proposes a fault diagnosis method. The third section presents a fault-tolerant operation strategy for triple-port hydrogen converter faults. The fourth section offers simulation results for verification. The fifth section compares the proposed method with existing solutions. The sixth section presents experimental validation using the RT-LAB platform.

## II. Operating Principle and Transmission Characteristics Analysis

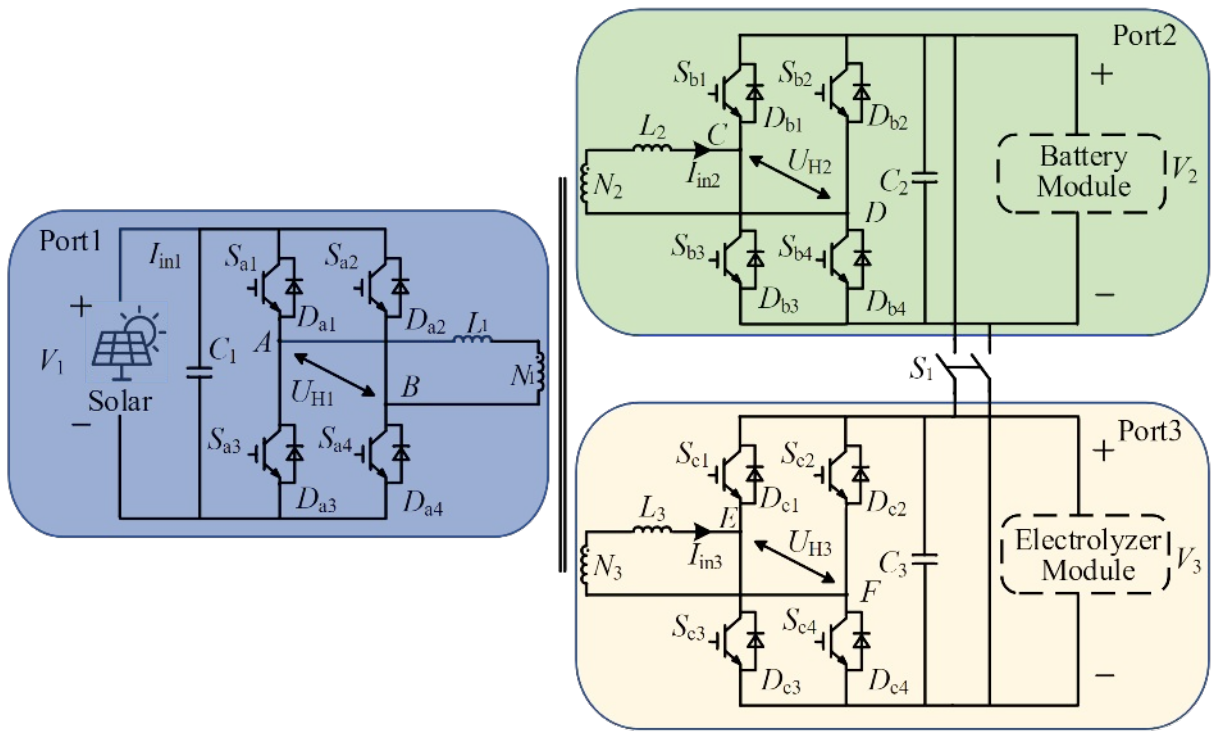

**Fig. 1.** Fault-tolerant Hydrogen Production Converter.

The topology of the triple-port hydrogen production converter is shown in Fig. 1. The electrolyzer module and battery module contain DC-DC converters. The three-winding transformer provides electrical isolation between the ports, while the full-bridge units enable bidirectional power flow. In Fig.1, the turns ratio of the three-winding transformer is $N_1$: $N_2$: $N_3$, and each full-bridge unit is composed of switching devices $S_{x1}$ to $S_{x4}$(x=a, b, c). The power transfer inductances $L_1$, $L_2$, and $L_3$ are formed by the leakage inductance of each port in the transformer, combined with auxiliary inductances. The voltages at the full-bridge unit ports are $U_{H1}$, $U_{H2}$, and $U_{H3}$, respectively, while the DC-side voltage stabilization capacitors at each port are $C_1$, $C_2$, and $C_3$.

According to the design requirements of the hydrogen production system, Port 1 of the triple-port hydrogen production converter is connected to a photovoltaic generation unit, Port 2 is connected to a battery storage unit, and Port 3 is connected to the electrolyzer unit. In the event of a switch fault, the faulty switch at the affected port can be locked, allowing power from the non-faulty ports to flow to the load unit of the faulty port via auxiliary switch $S_1$, ensuring stable system operation. Additionally, the power generated at Port 2 and Port 3 can be redirected as needed through auxiliary switch $S_1$, enabling power sharing between the ports.

### *A. Multi-Phase Shift Control for Triple-Port Converter*

The triple-port hydrogen production converter discussed in this paper employs Multi-Phase Shift (MPS) control, as illustrated in Fig. 2. MPS control introduces internal phase shift angles $\alpha_i$ (i=1,2,3) at each of the three ports of the TAB converter to regulate the duty cycle of the square-wave voltage $V_{Hi}$ on the transformer side. For simplicity, the transformer-side voltages $V_{Hi}$ for each port are depicted in Fig. 3, where the duty cycle of $V_{Hi}$ is defined as $d_i = (\pi - \alpha_i)/\pi$. Due to the periodic nature of the waveform, the voltage $V_{H1}$ at Port 1 is shifted left by $\alpha_{1/2}$, transforming the voltage waveform into an odd function. As shown in Fig. 3, the phase difference between the midpoints of the square-wave voltages $V_{Hi}$ at each port is defined as the external phase shift angle $\beta_{ij}$. $\beta'_{ij}$ represents the phase difference between the rising edges of the square-wave voltages output to the transformer side under the SPS (Single Phase Shift) control strategy.

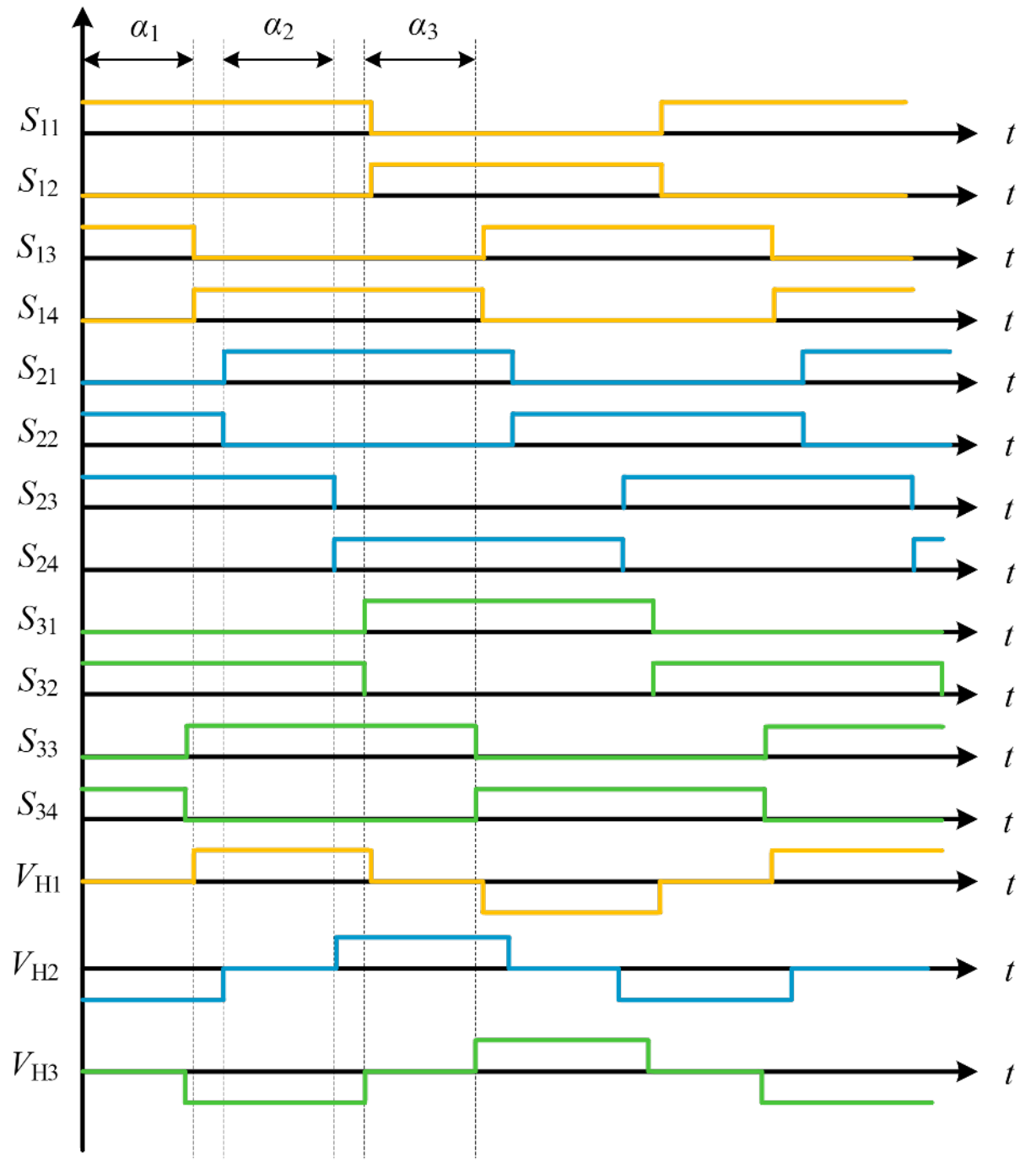

**Fig. 2.** Signal and square wave voltage of TAB converter.

## III. Fault Diagnosis and Fault Tolerance

### *A. Open-Circuit Fault Analysis*

In the triple-port hydrogen production converter, each port uses a full-bridge unit, making internal power switch device failures relatively common. When an open-circuit fault occurs in a power device, the corresponding anti-parallel diode can continue to operate normally, resulting in minimal changes in the circuit current.

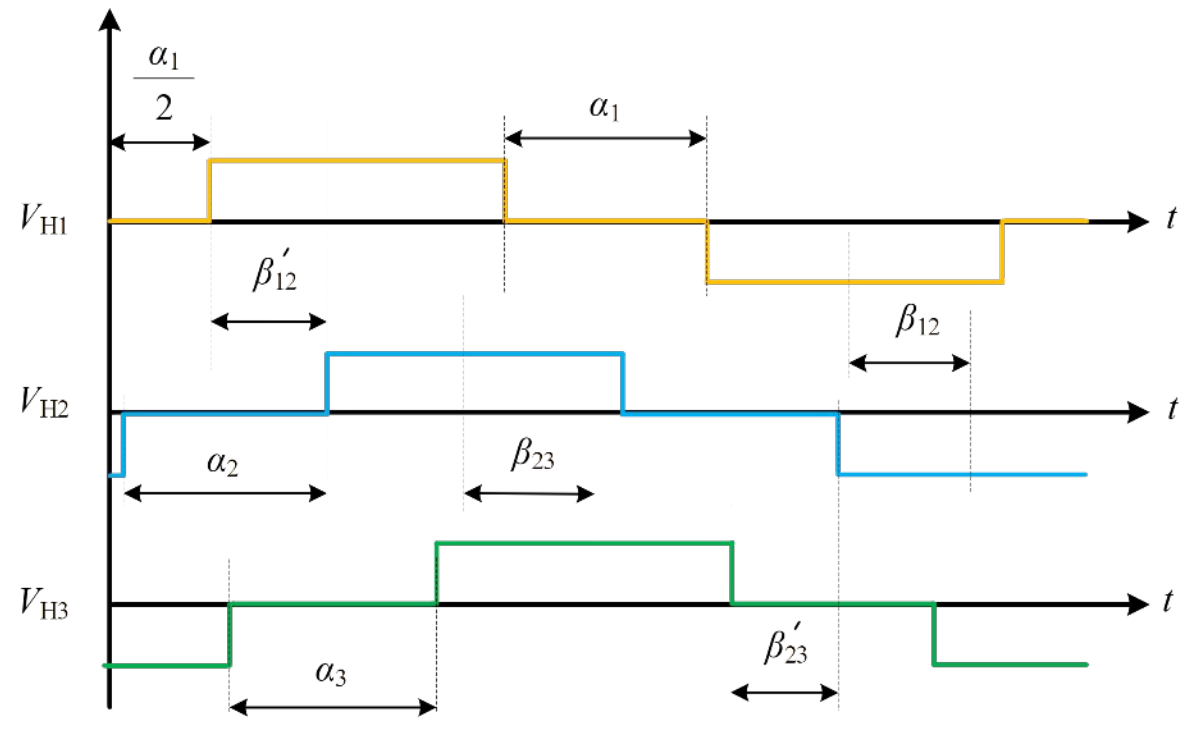

**Fig. 3.** Square wave voltage under MPS control.

As a result, the system can keep running even after an open-circuit fault occurs, which makes such faults difficult to detect. However, if the converter continues to operate in this faulty state, it may cause further damage to other internal components. Traditional fault diagnosis methods for triple-port converter typically require collecting voltage characteristic variables from all three ports and the midpoint voltages of the three bridges to perform logical judgments for fault diagnosis. This approach necessitates a large number of sensors and cannot diagnose multiple faults simultaneously. A detailed comparison is shown in TABLE I.

In this study, we collected 68,000 data points of input and output voltage and current from the three ports of the hydrogen production converter under various fault conditions using the RT-LAB platform. We employed recursive feature elimination for feature selection, followed by a component analysis of the data. Three key feature sets, each contributing more than 95% [26], were identified, as shown in Fig. 4.

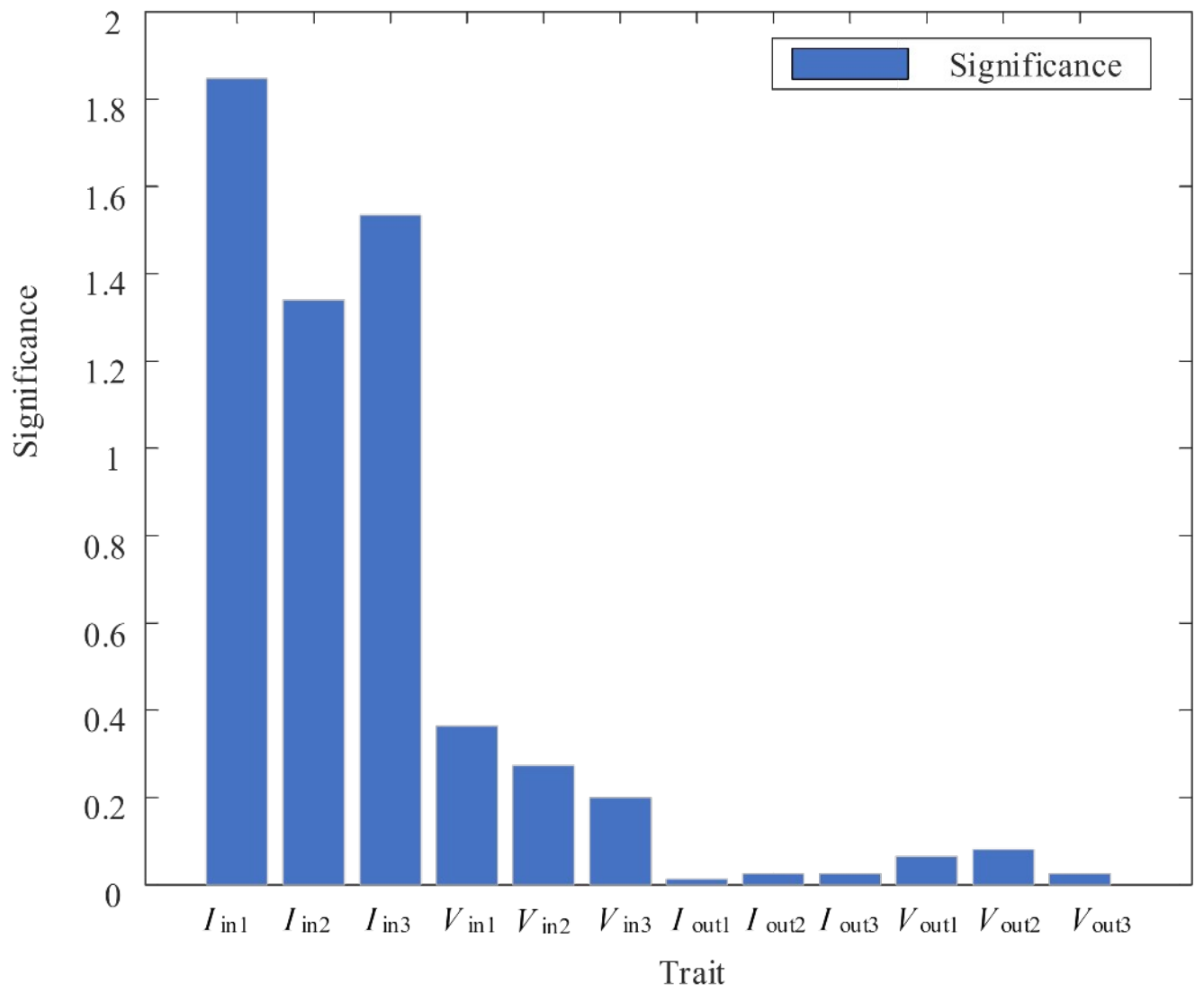

**Fig. 4.** Importance analysis.

These features correspond to the input current data for the three ports of the hydrogen production converter: $I_{in1}$, $I_{in2}$, and $I_{in3}$. Based on the importance analysis of these features, it was determined that the currents through the inductances $L_1$ at Port 1 ($I_{in1}$), $L_2$ at Port 2 ($I_{in2}$), and $L_3$ at Port 3 ($I_{in3}$) are sufficient to serve as characteristic signals for fault diagnosis.

Taking the open-circuit fault of the electrolyzer-side switches $S_{c1}$ and $S_{c2}$ as an example for analysis, Fig. 5 shows the waveforms of the AC voltage and inductance current after the open-circuit fault.

1) $t_0$-$t_1$: At this time, the inductance current $I_{L3}$ is greater than 0, and the current flows through $S_{c2}$ and $D_{c4}$, resulting in $V_{H3}$ being 0.
2) $t_1$-$t_2$: The inductance current $I_{L3}$ is less than 0, and $S_{c2}$ and $S_{c3}$ are conducting, resulting in $V_{H3}$ being equal to -$V_3$.
3) $t_2$-$t_3$: At time $t_2$, $S_{c1}$ and $S_{c2}$ experience a fault. The inductance current $I_{L3}$ is less than 0, and the current flows through $D_{c2}$ and $D_{c3}$, resulting in $V_{H3}$ being equal to -$V_3$.
4) $t_3$-$t_4$: At this time, $S_{c3}$ is conducting, the inductance current $I_{L3}$ is greater than 0, and the current flows through $D_{c1}$ and $S_{c3}$, resulting in $V_{H3}$ being 0.
5) $t_4$-$t_5$: With $S_{c1}$ and $S_{c3}$ faulty, only $S_{c4}$ is conducting. The inductance current $I_{L3}$ is greater than 0, and the current flows through $D_{c1}$ and $D_{c4}$, resulting in $V_{H3}$ being equal to $V_3$.

6) $t_5$-$t_6$: At this time, $S_{c2}$ and $S_{c4}$ are conducting, the inductance current $I_{L3}$ is greater than 0, and the current flows through $S_{c2}$ and $D_{c4}$, resulting in $V_{H3}$ being 0.

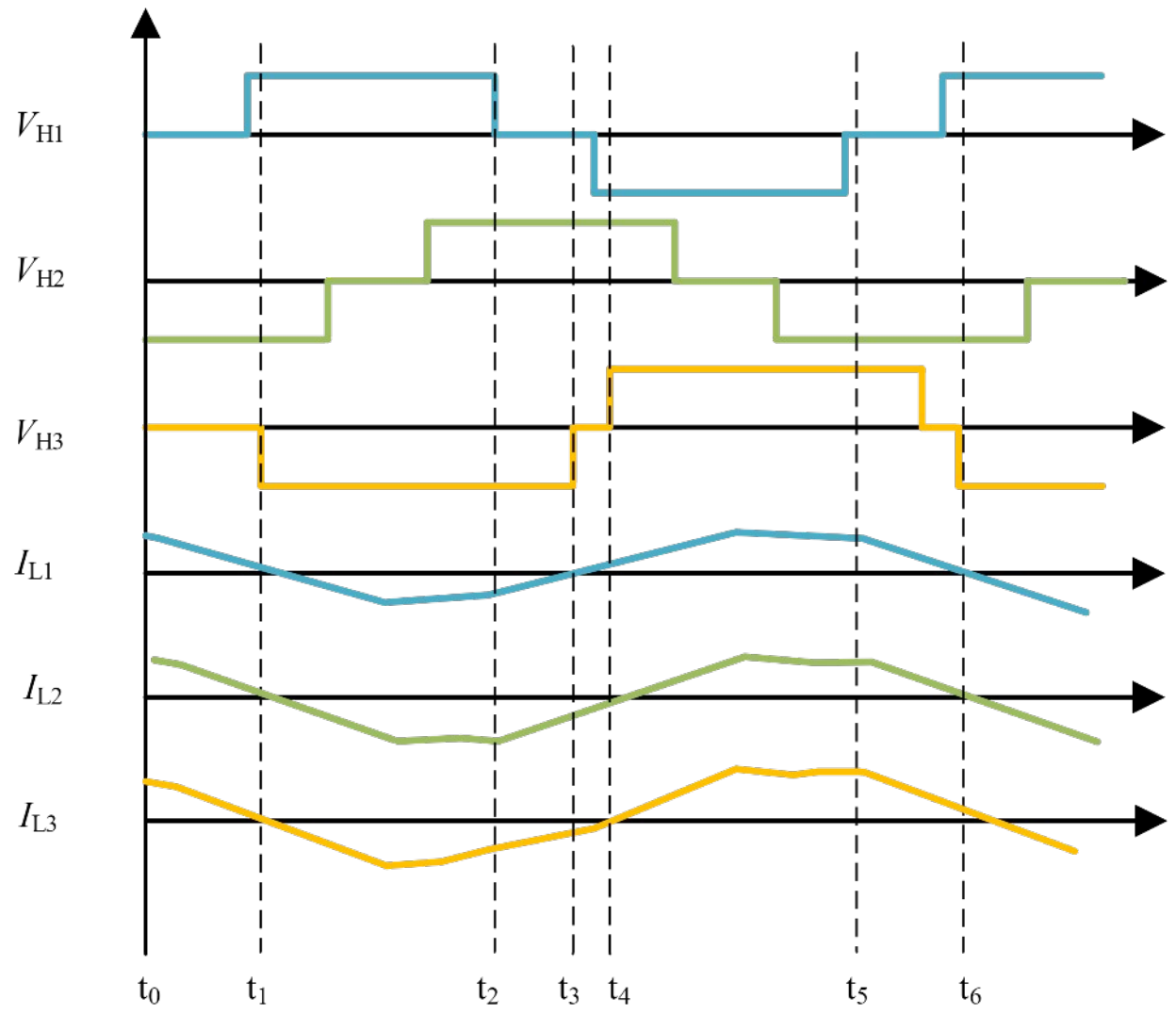

**Fig. 5.** $S_{c1}$ and $S_{c2}$ Open-Circuit Fault.

The triple-port hydrogen production converter is prone to various types of faults, and existing methods are often complex, making it difficult to directly locate faults based on the signals and incapable of diagnosing multiple faults simultaneously. The paper provides a theoretical analysis of multiple switch faults. Additionally, diagnosing single-switch faults in triple-port hydrogen converter typically requires more voltage and current sensors, which increases diagnostic costs. The diagnostic method proposed in this paper addresses these challenges by applying recursive feature elimination to the collected fault characteristic data, selecting the three features with the highest contribution, and then using a sparrow search algorithm-optimized random forest algorithm to process the selected data, enabling accurate fault type diagnosis.

### *B. Sparrow Search -Optimized Random Forest Algorithm*

The Sparrow Search Algorithm (SSA) is a novel swarm intelligence optimization algorithm known for its high search efficiency and simplicity in parameter settings. Random Forest (RF) is one of the most widely used machine learning methods, offering excellent predictive accuracy, the ability to quickly process large training datasets, and the capability to handle high-dimensional feature input samples without the need for dimensionality reduction. It is also suitable for applications with incomplete data. Therefore, this paper proposes a TAB fault diagnosis model based on a Sparrow Search Algorithm-Optimized Random Forest model.

The final classification decision of the Random Forest algorithm is:

$$H(x) = \arg\max_{Y} \sum_{i=1}^{K} I\left(h_i(x) = Y\right)$$

11* MERGEFORMAT ()

The Sparrow Optimization Algorithm mainly includes leader position updating, follower position updating, and predator avoidance position updating.

Leader position updating:

$$x_{new} = x_{best} - r_1 \left| x_{best} - x_{current} \right|$$

22* MERGEFORMAT ()

where $x_{best}$ is the current best position, $x_{current}$ is the current position of the sparrow, and $r_1$ is a random number within the range [0,1].

Follower position updating:

$$x_{new} = x_i + r_2\left(x_{leader} - x_i\right)$$

33* MERGEFORMAT ()

where $x_{leader}$ is the leader's position in the group, $x_{new}$ is the current position of the sparrow, and $r_2$ is a random number within the range [0,1].

Predator avoidance position updating:

$$x_{new} = x_{current} + r_3\left(x_{current} - x_{worst}\right)$$

44* MERGEFORMAT ()

where $x_{worst}$ is the worst position in the group, representing the possible location of a predator, and $r_3$ is a random number within the range [0,1]. The Sparrow Search Algorithm can be used to optimize and derive the optimal Random Forest model.

### *C. SSA-RF Fault Diagnosis Method*

The steps for fault diagnosis based on the SSA-RF method are as follows:

1) **Data acquisition**
   Training the Random Forest model requires a large amount of data. Through fault feature importance analysis, it is determined that the collected feature signals should be the input currents of the three ports. Three current sensors are used to collect the input current data from the three ports of the triple-port hydrogen production converter: $I_{in1}$, $I_{in2}$, and $I_{in3}$.
2) **Data preprocessing**
   Preprocessing the system data of the triple-port hydrogen production converter: The collected input current data $I_{in1}$, $I_{in2}$, and $I_{in3}$ are mapped to the range [0,1] and normalized to improve algorithm performance and ensure stable output results.
3) **Random Forest model training**
   Use 70% of the processed input current data Iin1, Iin2, and Iin3 for training and 30% for testing to build the Random Forest model [27].
4) **Sparrow Search Algorithm Optimization**
   Use Optimize the Random Forest model using the Sparrow Search Algorithm. First, initialize the population and the number of iterations, as well as the proportions of predators and joiners. Calculate the fitness of the initial population and rank them to determine the best and worst positions of the current sparrows. Calculate the fitness and update the positions of the sparrows. Evaluate whether the new positions are better; if they are, continue updating to find the optimal sparrow positions. If not, check if the maximum number of iterations has been reached. If the maximum number is reached, proceed to the next step;

otherwise, return to calculate the fitness of the initial population. Terminate the optimization algorithm and output the optimal parameters for the SR model. Input the optimal parameters into the SR model to test the diagnostic results and complete the construction of the optimal SSA-RF model.

5) **Optimized Random Forest Model Fault Diagnosis**
Use the Random Forest model: Test the actual fault diagnosis effectiveness using the optimized Random Forest model. The actual collected input current data $I_{in1}$, $I_{in2}$, and $I_{in3}$ from the three ports of the triple-port hydrogen production converter are processed in Step 2 and then input into the trained and optimized Random Forest network for identification. Accurately diagnose the fault locations of all switching devices $S_{a1}$-$S_{a4}$, $S_{b1}$-$S_{b4}$, and $S_{c1}$-$S_{c4}$ in the triple-port hydrogen production converter. Compare the network model's diagnostic output with the actual fault status of the switching devices in the triple-port hydrogen production converter to evaluate the network's performance.

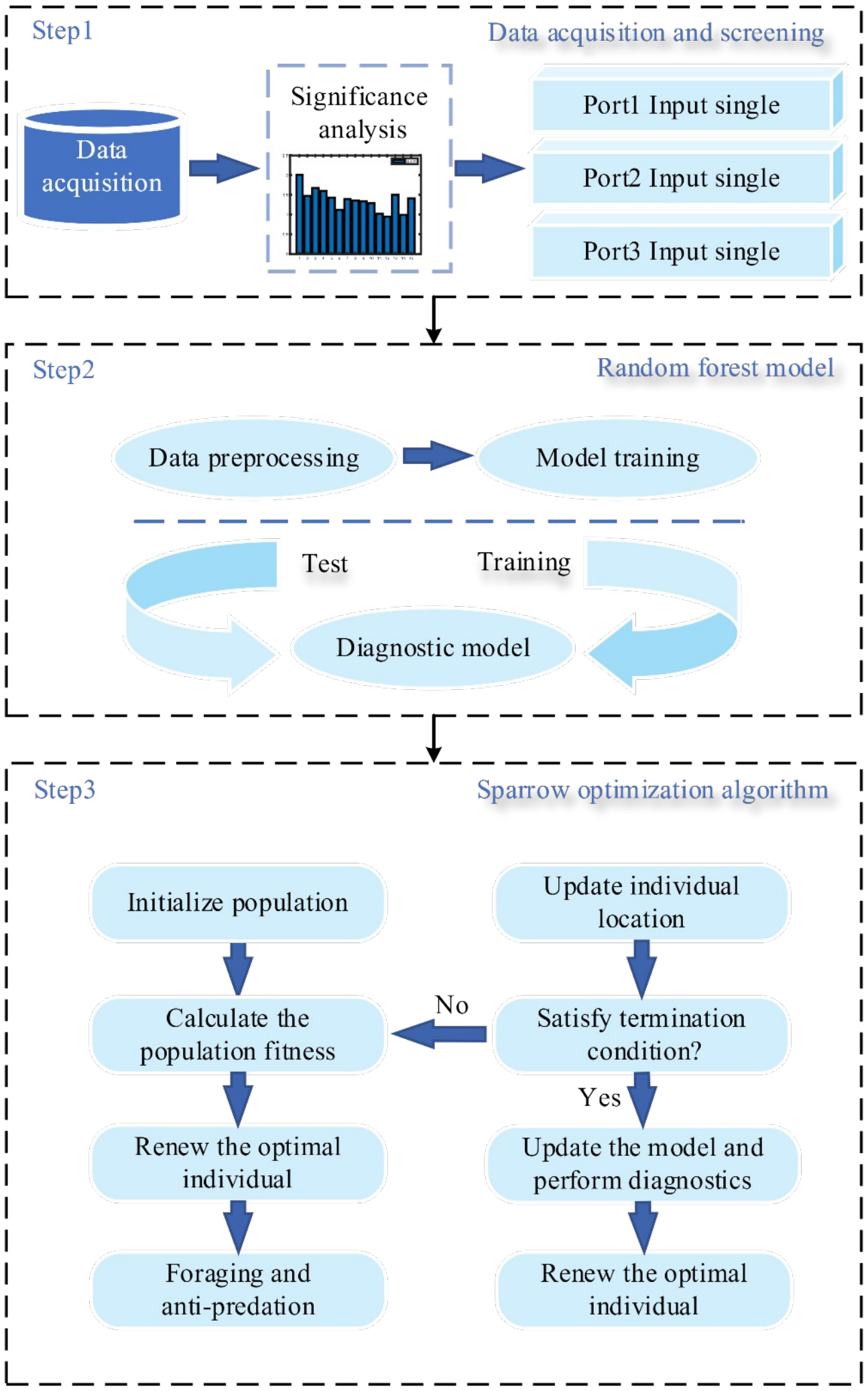

**Fig. 6.** Sparrow algorithm to optimize the flow chart of random forest model.

*D. Research on Fault-Tolerant Control Topology.*

When an open-circuit fault occurs in the power switching devices of a triple-port hydrogen production converter, it can lead to changes in the periodic average value of the port voltage and a bias in the inductor current. This, in turn, affects the stability of the circuit operation and can cause excessive heating of the triple-port transformer. To prevent these issues, a fault-tolerant control strategy based on topology reconfiguration has been designed.

The fault-tolerant topology proposed in this paper is shown in Fig. 7. Taking the normal operation of a triple-port hydrogen production converter as an example when a module fault occurs, the system can resume normal operation by quickly adjusting the auxiliary branch switch states.

To ensure the safe and stable operation of the entire system and to prevent further damage from an open-circuit fault in the power switching devices of the triple-port hydrogen production converter, an auxiliary switch branch is added to the converter topology. In the event of an open-circuit fault in the converter power switching devices, fault-tolerant operation is carried out by utilizing multiple switches between the battery unit at Port 2 and the electrolyzer unit at Port 3, enabling the system to continue normal operation.

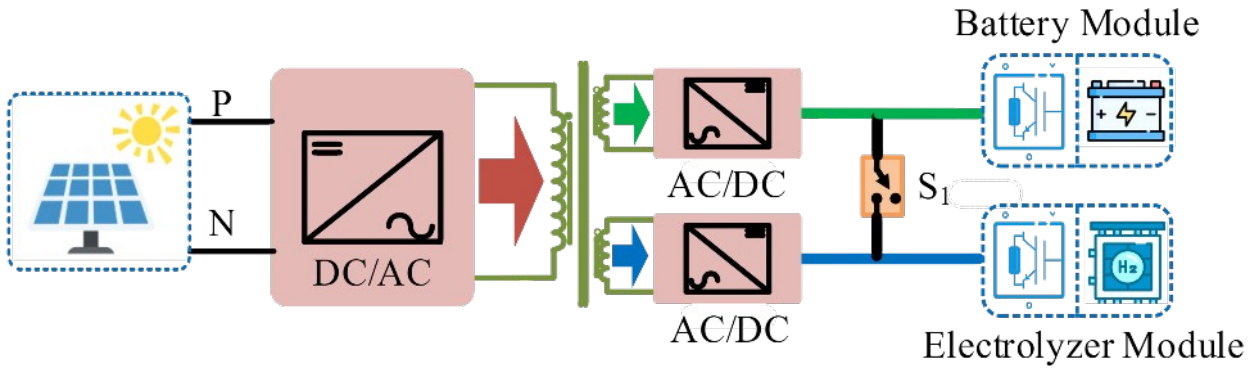

**Fig. 7.** Normal operating condition.

*E. Photovoltaic Side Fault-Tolerant Control*

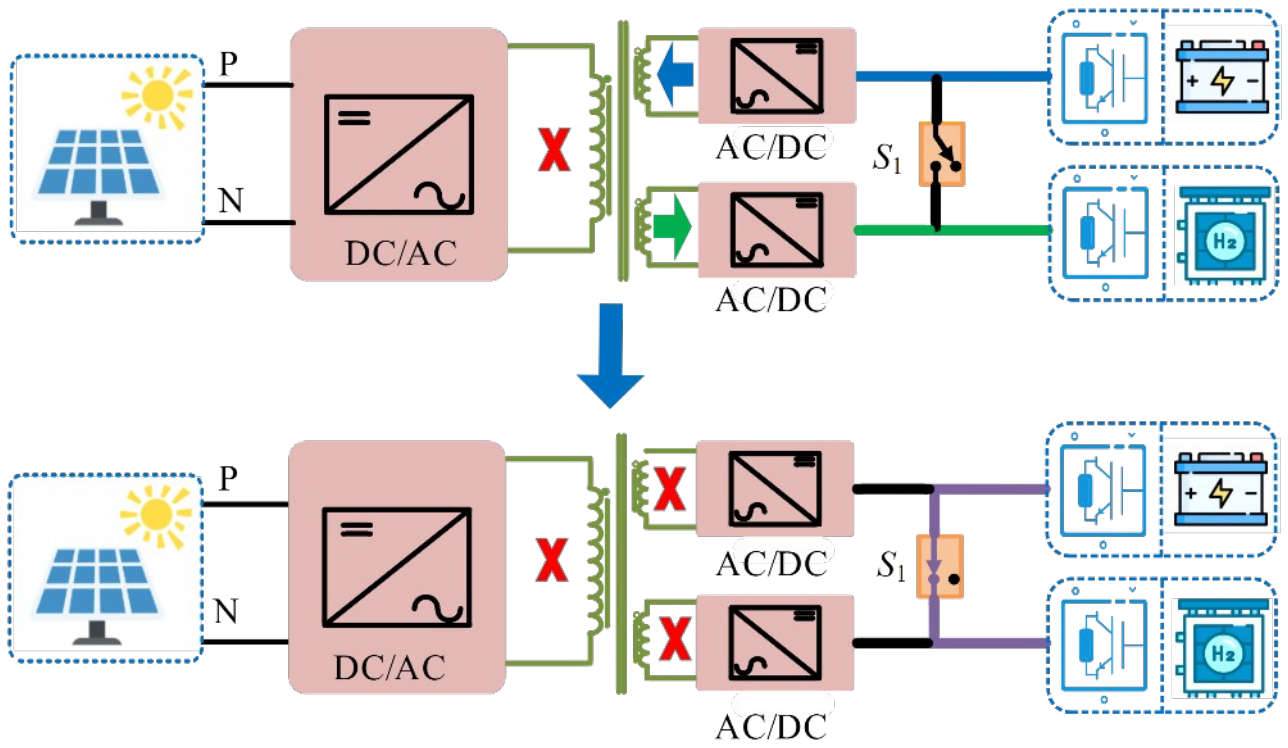

**Fig. 8.** Port one fault tolerance.

As shown in Fig. 8 when a fault occurs on the photovoltaic side, the system implements the following fault-tolerant control process: Ports 2 and 3 are locked, switch $S_1$ is closed, and the battery supplies power to the electrolyzer, ensuring its normal operation. If two or three port-switching devices fail, the system can lock all three ports and rely solely on the battery to power the electrolyzer. As shown in Fig. 9(a), from 0 to $t_1$, the system operates normally. Between $t_1$ and $t_2$, a fault occurs in the switching device at Port 1. During this period, $I_{out2}$ is slightly negative, $I_{out3}$ is slightly positive, and the power on the electrolyzer side is slightly above zero, with a small amount of energy flowing from the battery to the hydrogen production electrolyzer through the transformer windings.

At $t_2$, fault-tolerant control is activated, the auxiliary switch operates, and the battery supplies power to the electrolyzer, restoring the electrolyzer side to its normal energy level.

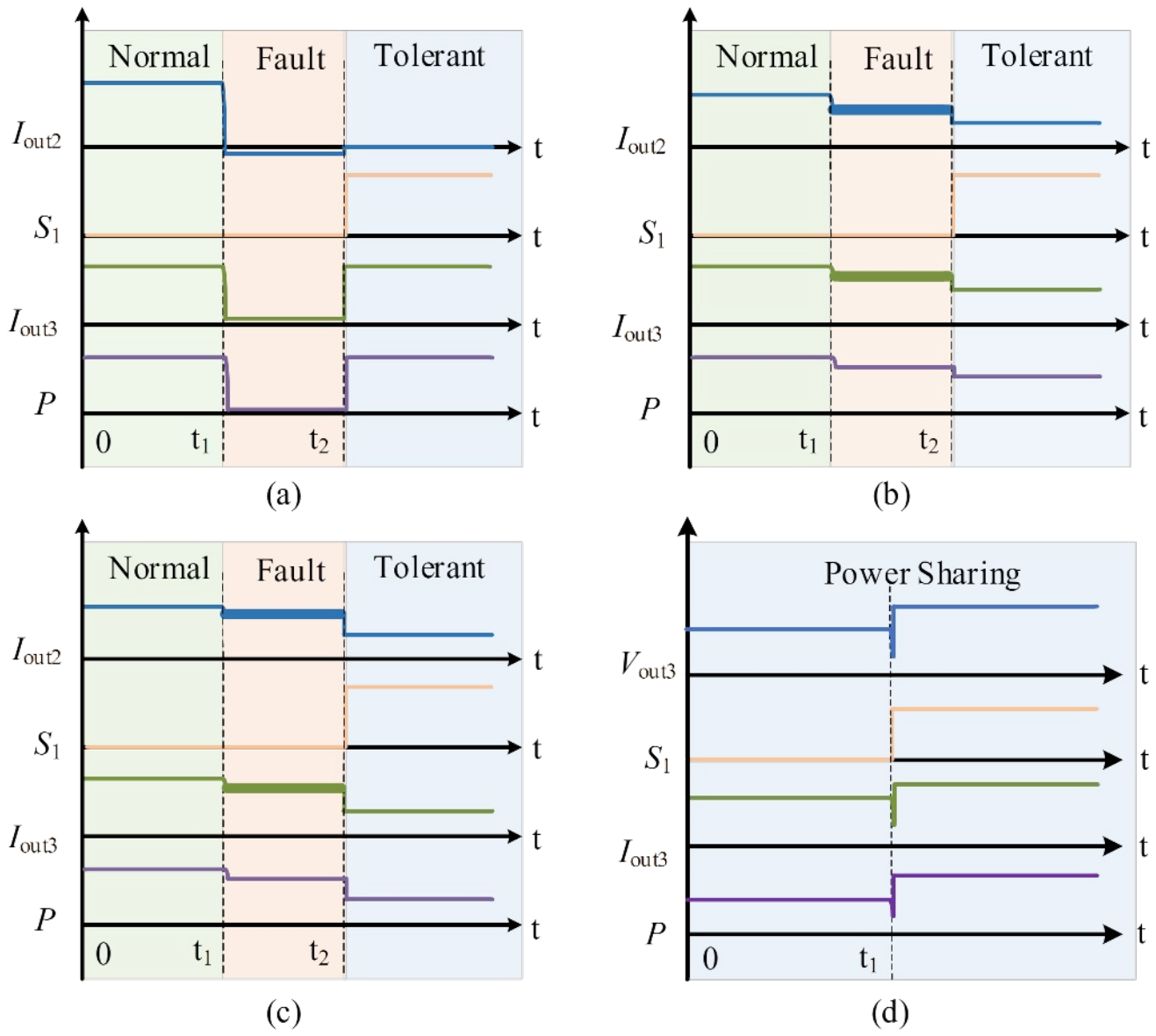

**Fig. 9.** Fault tolerant analysis (a)Port one fault tolerance. (b) Port two fault tolerance.(c) Port three fault tolerance. (d) Power sharing process.

### F. Battery Side Fault-Tolerant Control

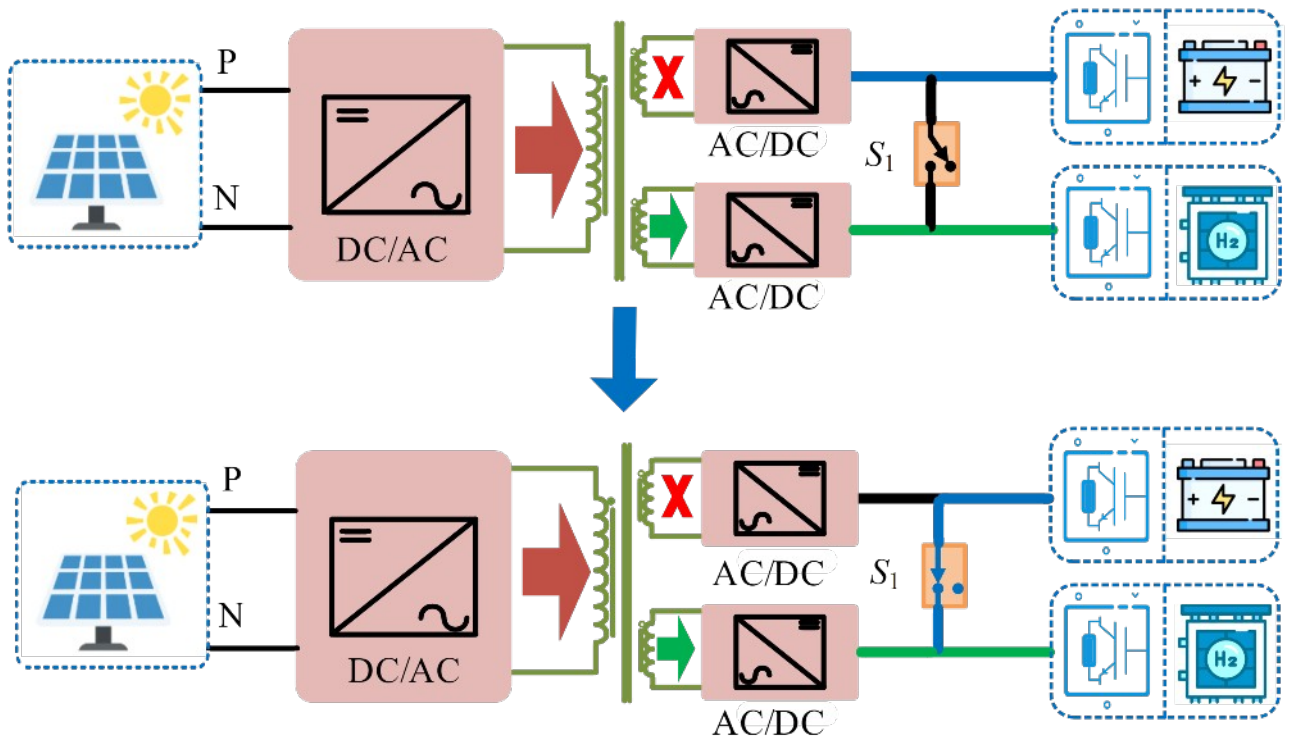

**Fig. 10.** Port two fault tolerance.

As shown in Fig. 10, when a fault occurs on the battery side, the system implements the following fault-tolerant control process: Close switch $S_1$, allowing the photovoltaic side to supply power to both the battery and the electrolyzer through Port 3, ensuring normal system operation. As shown in Fig. 9(b), from 0 to $t_1$, the system operates normally. Between $t_1$ and $t_2$, a fault occurs in the switching device at Port 2. During this period, $I_{out2}$ and $I_{out3}$ are unable to output stable DC current. At $t_2$, fault-tolerant control is activated, and the auxiliary switch operates, allowing Port 3 to simultaneously supply power to both the electrolyzer and the battery.

### G. Electrolyzer Side Fault-Tolerant Control

As shown in Fig. 11, when a fault occurs on the electrolyzer side, the system implements the following fault-tolerant control process: Close switch $S_1$, allowing the photovoltaic side to supply power to both the battery and the electrolyzer through Port 2, ensuring normal system operation. As shown Fig. 9(c), from 0 to $t_1$, the system operates normally. Between t1 and $t_2$, a fault occurs in the switching device at Port 3. During this period, $I_{out2}$ and $I_{out3}$ are unable to output stable DC current. At $t_2$, fault-tolerant control is activated, and the auxiliary switch operates, allowing Port 2 to simultaneously supply power to both the electrolyzer and the battery.

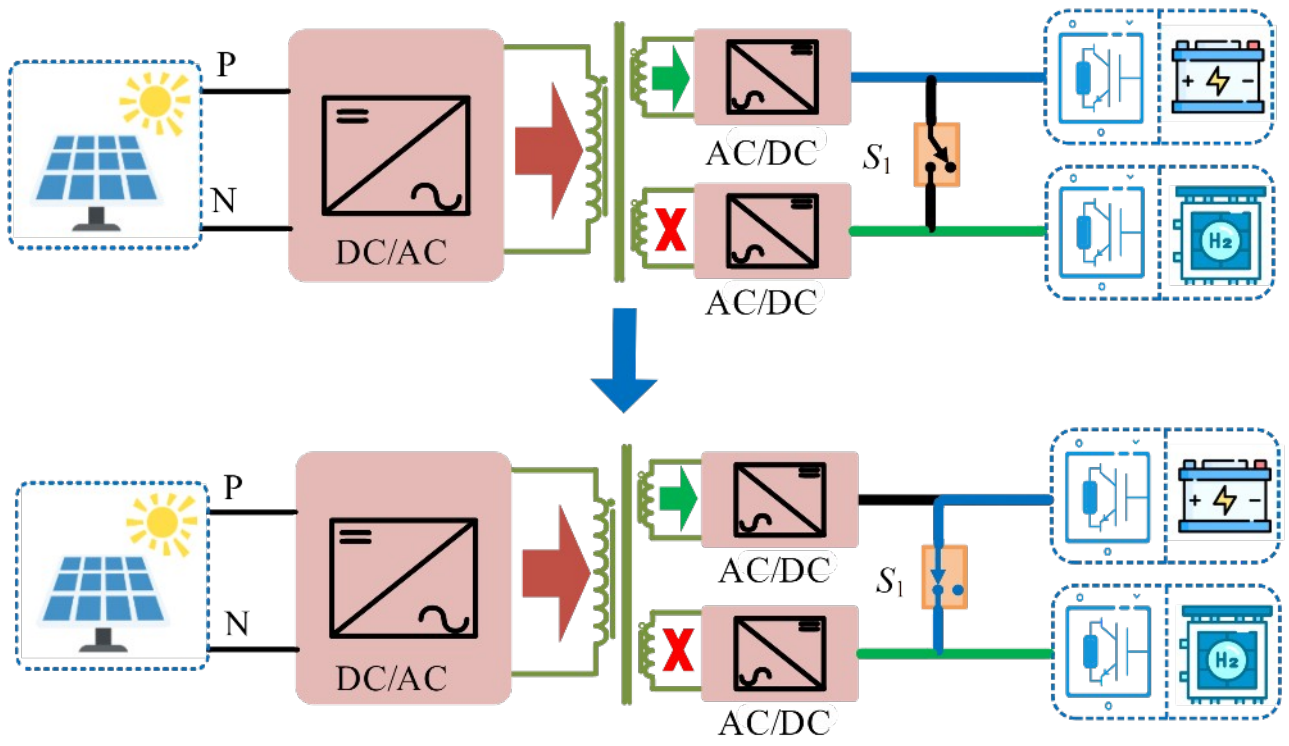

**Fig. 11.** Port three fault tolerance.

### H. Power Sharing

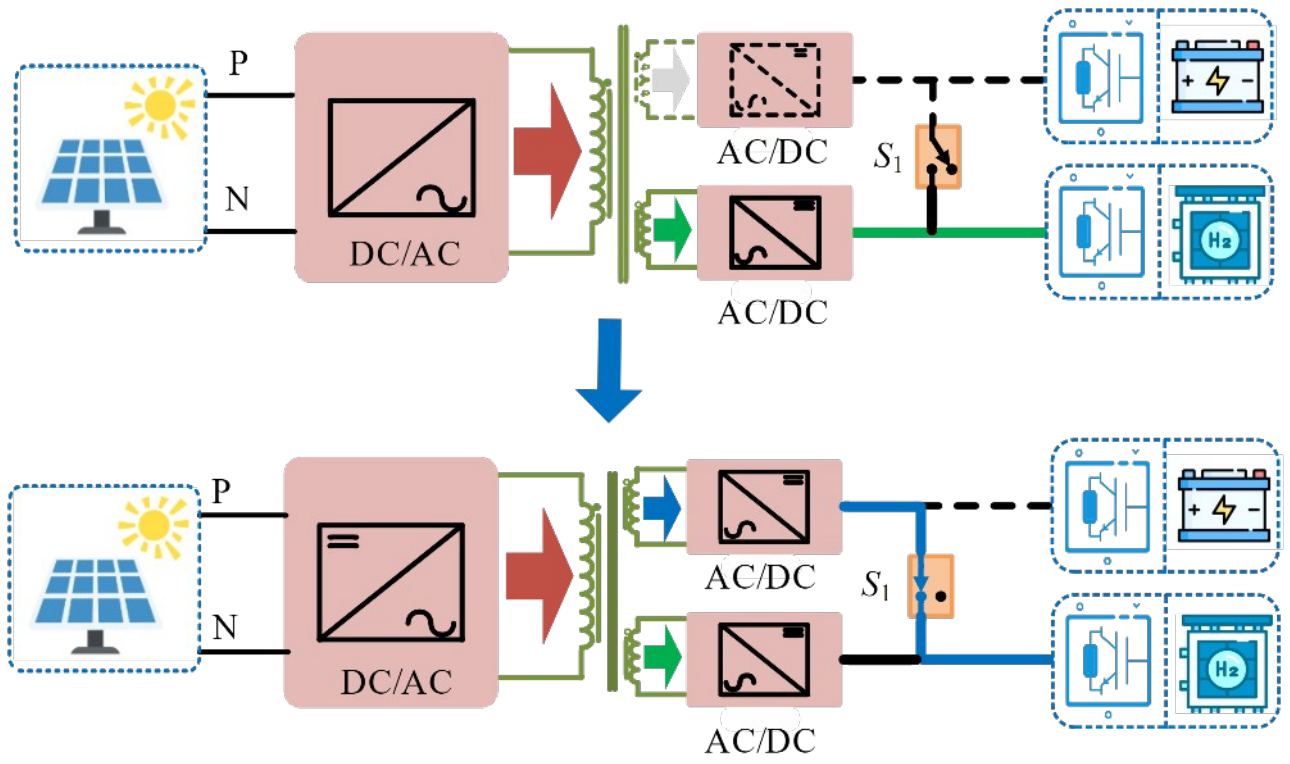

**Fig. 12.** Power sharing process.

As shown in Fig. 12 and Fig. 9(d), when the battery does not require charging, switch $S_1$ can be closed as needed, allowing ports 2 and 3 to simultaneously supply power to the electrolyzer, thereby increasing the power available to the electrolyzer. From 0 to $t_1$, the switch remains inactive. When the electrolyzer side requires a higher power input, the auxiliary switch activates, enabling both Port 2 and Port 3 to supply power to the electrolyzer simultaneously, thus increasing the electrolyzer's power and achieving power sharing.

## IV. Simulation and Experiment

### A. Fault diagnosis simulation verification

In this study, 25000 input current data samples $I_{in1}$, $I_{in2}$, and $I_{in3}$ were taken for each fault scenario involving switching devices $S_{a1}$-$S_{a4}$, $S_{b1}$-$S_{b4}$, and $S_{c1}$-$S_{c4}$, respectively, for training and testing. 70% of the data was used to train the Random Forest model, and 30% was used for diagnostic testing [18]. The diagnostic test results are shown in Fig. 13.

In Fig. 13, the predicted results 1-12 correspond to the 12 fault types associated with the failure of switching devices $S_{a1}$-$S_{a4}$, $S_{b1}$-$S_{b4}$, and $S_{c1}$-$S_{c4}$. To validate the model's ability to diagnose double-switch faults, four double-switch fault types were tested: 2.5 represents simultaneous faults in $S_{a1}$ and $S_{a4}$, 4.5 represents simultaneous faults in $S_{a4}$ and $S_{b1}$, 6.5 represents simultaneous faults in $S_{a3}$ and $S_{c4}$, and 9.5 represents simultaneous faults in $S_{c1}$ and $S_{c2}$. The overall fault diagnosis accuracy of the test set reached 99.7451%.

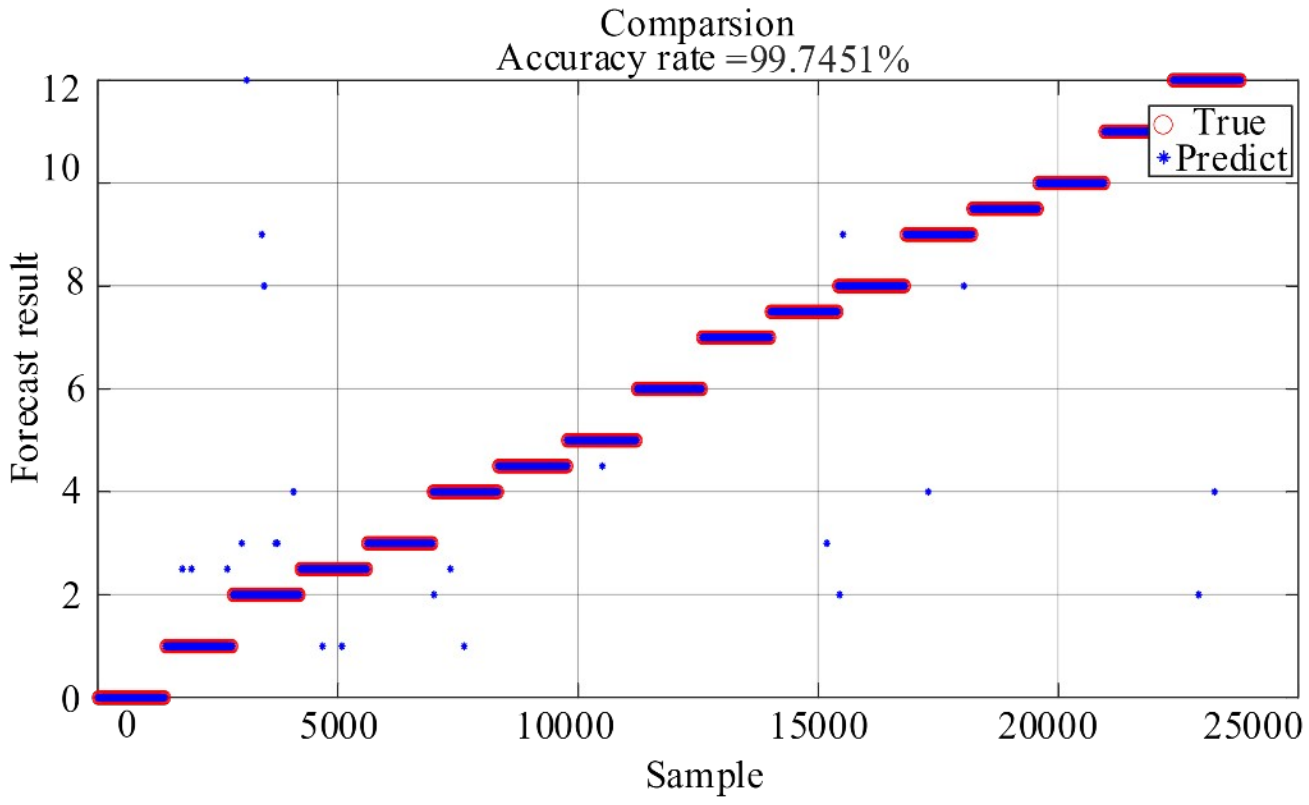

**Fig. 13.** SSA-RF Model test set diagnostic results.

*B. Fault tolerant control simulation verification*

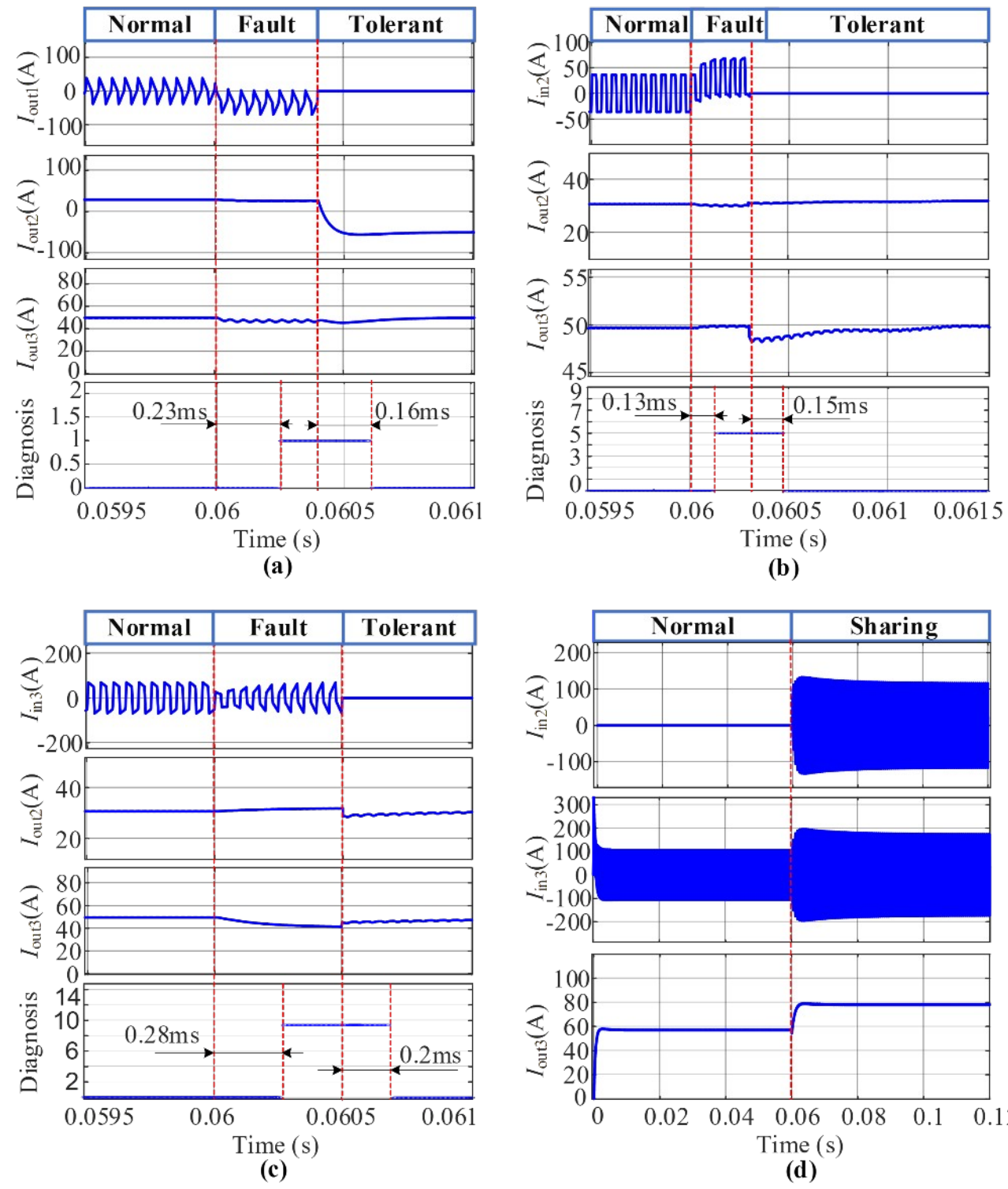

**Fig. 14.** Fault simulation (a) Port one fault simulation. (b) Port two fault simulation. (c) Port three fault simulation. (d) Power sharing simulation.

Simulation parameters are shown in TABLE III Experimental AND Simulation Parameters. As shown in Fig. 14(a), an open-circuit fault occurs in the full-bridge unit of Port 1 at 0.06 s. After 0.23 ms, fault diagnosis is performed, and a fault signal is output. At 0.0604 s, the system performs fault-tolerant operations, and 0.16 ms later, a normal operation signal is output. Port 2 and Port 3 are locked, and only the battery is used to supply power to the electrolyzer unit, ensuring that the electrolyzer unit operates under normal conditions. This enhances the stability of the hydrogen production system, demonstrating the effectiveness of the proposed diagnosis and fault-tolerant scheme.

As shown in Fig. 14(b), an open-circuit fault occurs in the full-bridge unit of Port 2 at 0.06 s. After 0.13ms, fault diagnosis is performed, and a fault signal is output. At 0.0603 s, the system performs fault-tolerant operations, and 0.15 ms later, a normal operation signal is output. Port 2 is locked, and the full-bridge unit of Port 3 supplies power to both the battery and the electrolyzer units, ensuring that the electrolyzer unit operates under normal conditions. This enhances the stability of the hydrogen production system, further validating the effectiveness of the proposed diagnosis and fault-tolerant scheme.

As shown in Fig. 14(c), an open-circuit fault occurs simultaneously in $S_{c1}$ and $S_{c2}$ within the full-bridge unit of Port 3 at 0.06 s. After 0.28 ms, fault diagnosis is performed, and a fault signal is output. At 0.0605 s, the system performs fault-tolerant operations, and 0.2ms later, a normal operation signal is output. Port 3 is locked, and the full-bridge unit of Port 2 supplies power to both the battery and the electrolyzer units, ensuring that the electrolyzer unit operates under normal conditions. This enhances the stability of the hydrogen production system, proving the effectiveness of the proposed diagnosis and fault-tolerant scheme.

As shown in Fig. 14(d), the power-sharing simulation waveform illustrates that when the battery is fully charged and the electrolyzer side requires more power, the auxiliary switch can be activated to allow both Port 2 and Port 3 to simultaneously supply power to the electrolyzer, increase the power of the electrolyzer achieving power sharing.

## V. Scheme Comparison

As shown in TABLE I and TABLE II, compared to existing methods such as those in [17], [18], [19], [20], and [21], which can only diagnose single-switch faults, this paper employs intelligent algorithms to diagnose multiple-switch faults. Compared to the requirement of four voltage sensors in [19] and three current sensors in [20] for diagnosing faults in the DAB converter, as well as the three voltage sensors and two current sensors needed in [17] for diagnosing faults in the TAB converter, the fault diagnosis method proposed in this paper requires only three current sensors to diagnose faults in a TAB converter. Compared to existing TAB diagnosis methods, such as in [17], which takes 10 signal cycles ($T_s$) to complete fault diagnosis, the method proposed in this paper only requires only a range of 2Ts to 2.5Ts, making it more efficient while meeting the safety, continuity, and efficiency requirements for green hydrogen production.

Regarding fault-tolerant control strategies, compared to existing methods such as those in [17], [19], and [23], which reduce the input power on the load side after fault-tolerant control, the method proposed in this paper ensures that the load-side power remains stable.

Additionally, existing methods, including the proposals mentioned in [17], [19], [23], [24], and [25], are unable to achieve power-sharing between converter ports. The fault-tolerant control strategy proposed in this paper can realize power sharing.

TABLE I Comparison of Diagnostic Schemes

| No. | Application topology | Fault location | Number of sensors | Multi-tube fault diagnosis | Diagnostic speed |
|---|---|---|---|---|---|
| [18] | H-Bridge Rectifiers | Yes | One voltage sensor and one current sensor | No | $2T_s$ |
| [21] | Inverter | Yes | One voltage sensor and one current sensor | No | $2T_s$ |
| [19] | DAB | Yes | Four voltage sensors | No | $1T_s$ |
| [20] | DAB | Yes | Three current sensors | No | $4T_s$ |
| [17] | TAB | Yes | Three voltage sensors and two current sensors | No | $10T_s$ |
| This paper | TAB | Yes | Three current sensors | Yes | Range of $2T_s$ to $2.5T_s$ |

TABLE II Comparison of Fault-Tolerant Control Schemes

| No. | Application topology | Fault-tolerant method | Degree of difficulty | Power sharing | Restrictions for TAB |
|---|---|---|---|---|---|
| [19] | DAB | Control strategy | High | No | Power transmission capacity will be reduced |
| [17] | TAB | Control strategy | Medium | No | The load side power is only 75% of the original |
| [23] | DAB | Control strategy | Medium | No | There will be low-frequency oscillations |
| [24] | DAB | Control strategy and blocking capacitors | High | No | Not suitable for TAB |
| [25] | DC-DC converter | Control strategy | High | No | Too complex for TAB |
| This paper | TAB | Control strategy and new topological structures | Low | Yes | Simple and fast; Suitable for TAB |

## VI. Experimental Verification

To validate the fault diagnosis and fault-tolerant operation strategies proposed in this paper, experimental verification was conducted on the TAB fault diagnosis and fault-tolerant control. The platform is shown in Fig. 15, and the parameters of the experimental platform are shown in TABLE III. The results confirm the correctness and effectiveness of the proposed fault diagnosis and fault-tolerant strategies.

### *A. Port one fault diagnosis and fault tolerance experiment*

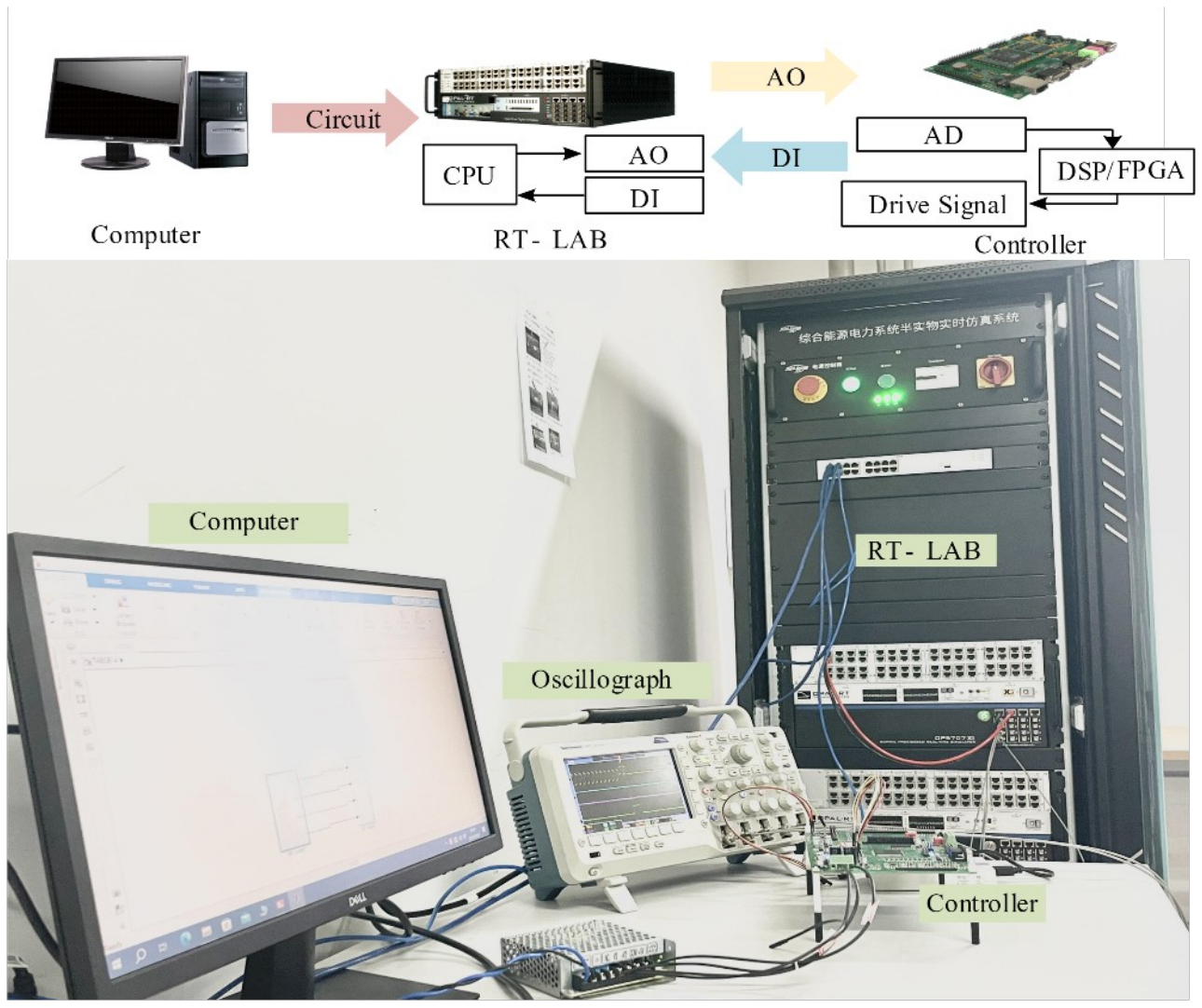

**Fig. 15.** Experimental platform.

In Fig. 16, the experimental waveform demonstrates fault diagnosis and fault-tolerant control when an open-circuit fault occurs in the switch Sa1 of Port 1 in the triple-port hydrogen production converter. At time t0, Sa1 experiences an open-circuit fault, causing a current bias at Port 1 and a reduction in the current flowing into the electrolyzer unit. After 0.32 ms, at time $t_1$, a fault diagnosis signal with a value of 1 is output. 0.06 ms later, at $t_2$, fault-tolerant control is initiated, locking the full-bridge units of Ports 2 and 3. After 0.24 ms, at time $t_3$, the fault signal is cleared, and the current flowing into the electrolyzer unit returns to normal, ensuring the electrolyzer unit operates under normal conditions.

TABLE III Experimental And Simulation Parameters

| Argument | Value | Argument | Value |
|---|---|---|---|
| Input voltage | 144V | $L_1$ | 100μH |
| Battery voltage | 24V | $L_2$ | 100μH |
| Load voltage | 24V | $L_3$ | 160μH |
| Transformer ratio | 6:1:1 | Filter inductance | 100μH |
| Power | 10kW | Filter capacitance | 470μH |
| Excitation inductance | 30mH | Switching frequency | 10kHz |

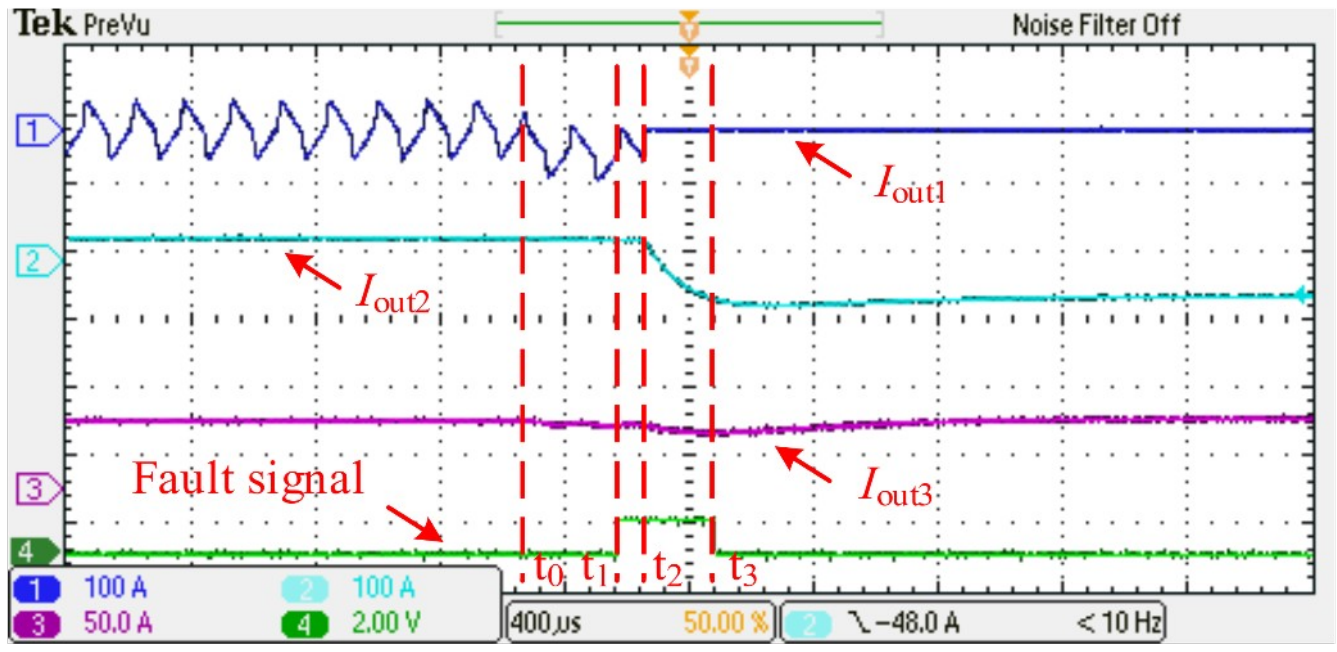

**Fig. 16.** $S_{a1}$ fault experiment.

In Fig. 17, at time $t_0$, $S_{b1}$ experiences an open-circuit fault, causing a current bias at Port 2 and reducing the current flowing into the battery unit. After 0.24 ms, at time $t_1$, a fault diagnosis signal with a value of 5 is output. 0.05 ms later, at time $t_2$, fault-tolerant control is initiated, locking Port 2, and the full-bridge unit of Port 3 supplies power to both the battery and electrolyzer units. After 0.23 ms, at time $t_3$, the fault signal is cleared, the current flowing into the battery unit returns to normal, and the current flowing into the electrolyzer unit remains unaffected, proving the effectiveness of the proposed diagnosis and fault-tolerant scheme.

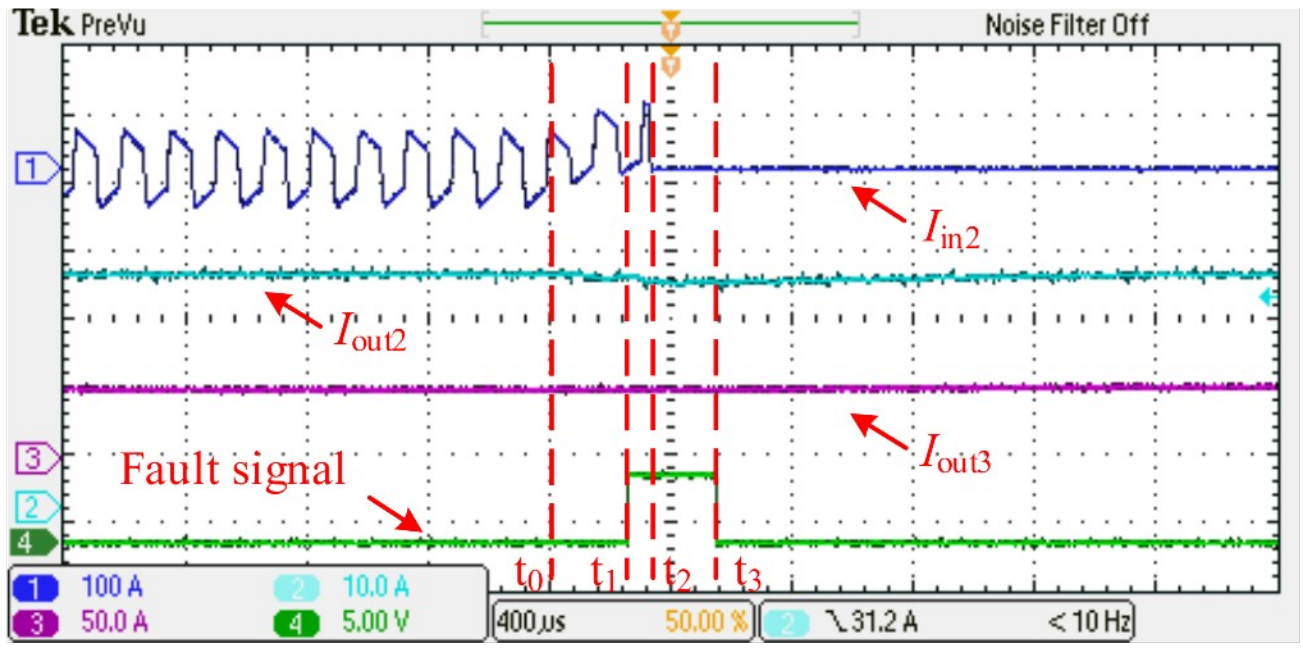

**Fig. 17.** $S_{b1}$ fault experiment.

In Fig. 18, at time $t_0$, $S_{c2}$ experiences an open-circuit fault, causing a current bias at Port 3. After 0.19 ms, at time $t_1$, a fault diagnosis signal with a value of 10 is output. 0.08 ms later, at time $t_2$, fault-tolerant control is initiated, locking Port 3, and the full-bridge unit of Port 2 supplies power to both the battery and electrolyzer units. After 0.18 ms, at time $t_3$, the fault signal is cleared, the current flowing into the electrolyzer unit returns to normal, and the current flowing into the electrolyzer unit remains unaffected, proving the effectiveness of the proposed diagnosis and fault-tolerant scheme.

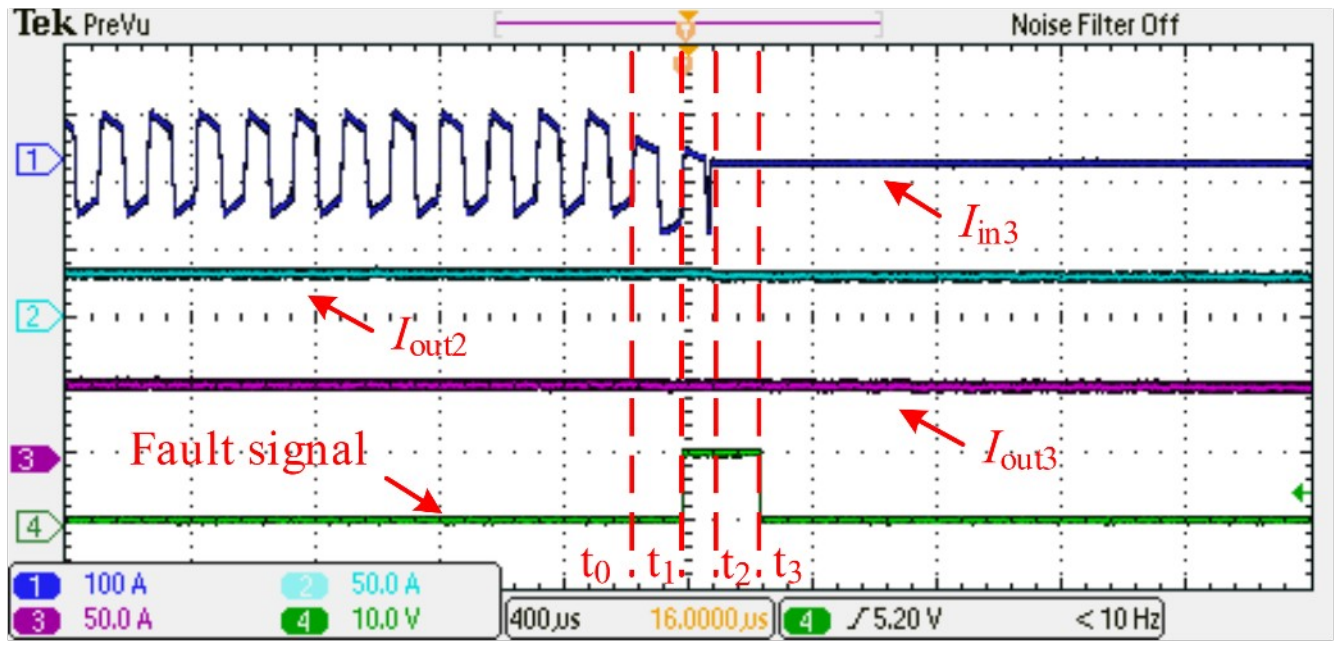

**Fig. 18.** $S_{c2}$ fault experiment.

*B. Multi-switch fault diagnosis and fault tolerance*

In Fig. 19, the experimental waveform demonstrates fault diagnosis and fault-tolerant control when an open-circuit fault occurs simultaneously in the switches $S_{c1}$ and $S_{c2}$ of Port 3 in the triple-port hydrogen production converter.

At time $t_0$, $S_{c1}$ and $S_{c2}$ experience open-circuit faults, causing a current bias at Port 3 and reducing the current amplitude. The current flowing into the electrolyzer unit decreases. After 0.25 ms, at time $t_1$, a fault diagnosis signal with a value of 9.5 is output. 0.05 ms later, at time $t_2$, fault-tolerant control is initiated, locking Port 3, and the full-bridge unit of Port 2 supplies power to both the battery and electrolyzer units. After 0.12 ms, at time $t_3$, the fault signal is cleared, the current flowing into the electrolyzer unit returns to normal, and the current flowing into the battery unit remains unaffected, demonstrating the effectiveness of the proposed diagnosis and fault-tolerant scheme.

In Fig. 20, at time $t_0$, $S_{c1}$ and $S_{c4}$ experience open-circuit faults, causing a current bias at Port 3 and reducing the current amplitude. The current flowing into the electrolyzer unit decreases. After 0.21 ms, at time $t_1$, a fault diagnosis signal with a value of 10.5 is output. 0.07 ms later, at time $t_2$, fault-tolerant control is initiated, locking Port 3, and the full-bridge unit of Port 2 supplies power to both the battery and electrolyzer units. After 0.11 ms, at time $t_3$, the fault signal is cleared, the current flowing into the electrolyzer unit returns to normal, and the current flowing into the battery unit remains unaffected, demonstrating the effectiveness of the proposed diagnosis and fault-tolerant scheme.

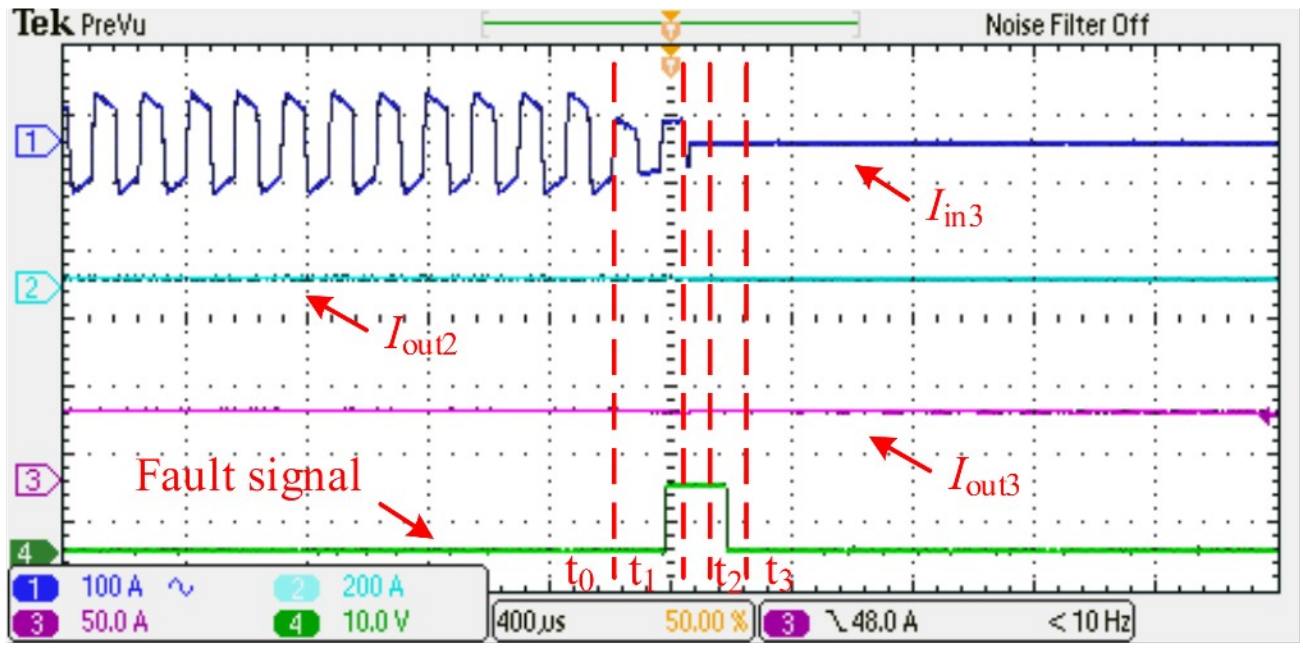

**Fig. 19.** $S_{c1}$ and $S_{c2}$ fault experiment.

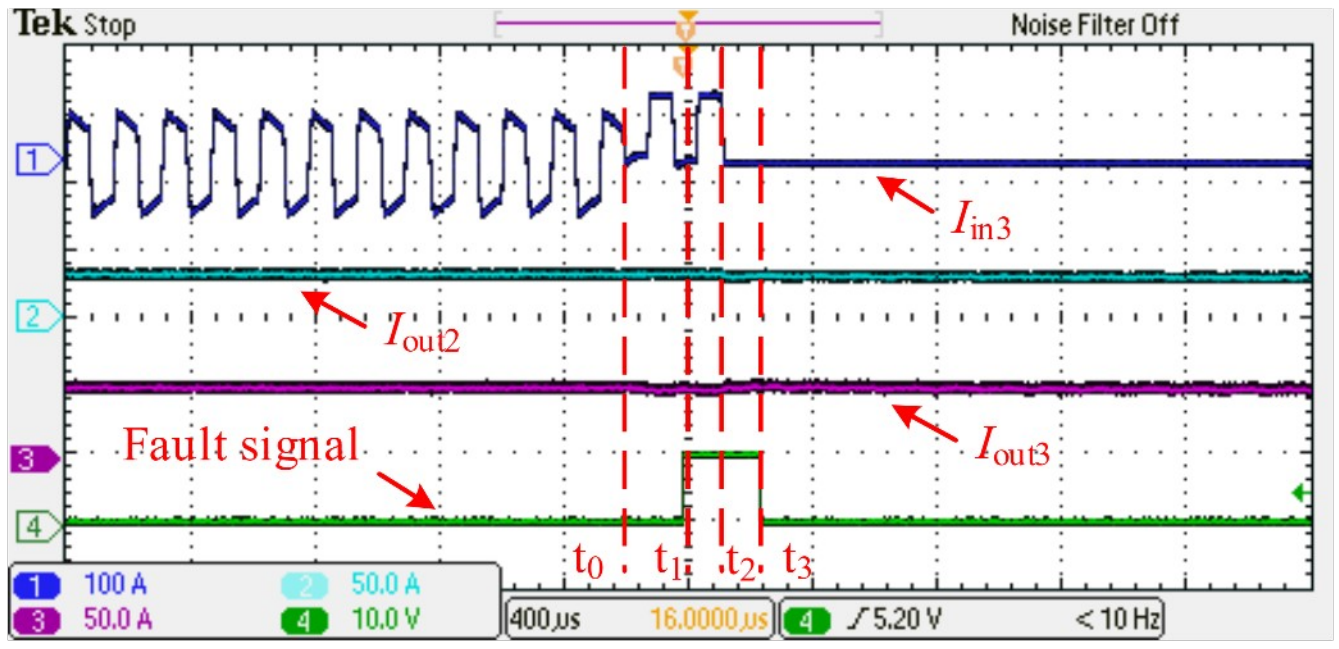

**Fig. 20.** $S_{c1}$ and $S_{c4}$ fault experiment.

*C. Power sharing experiment*

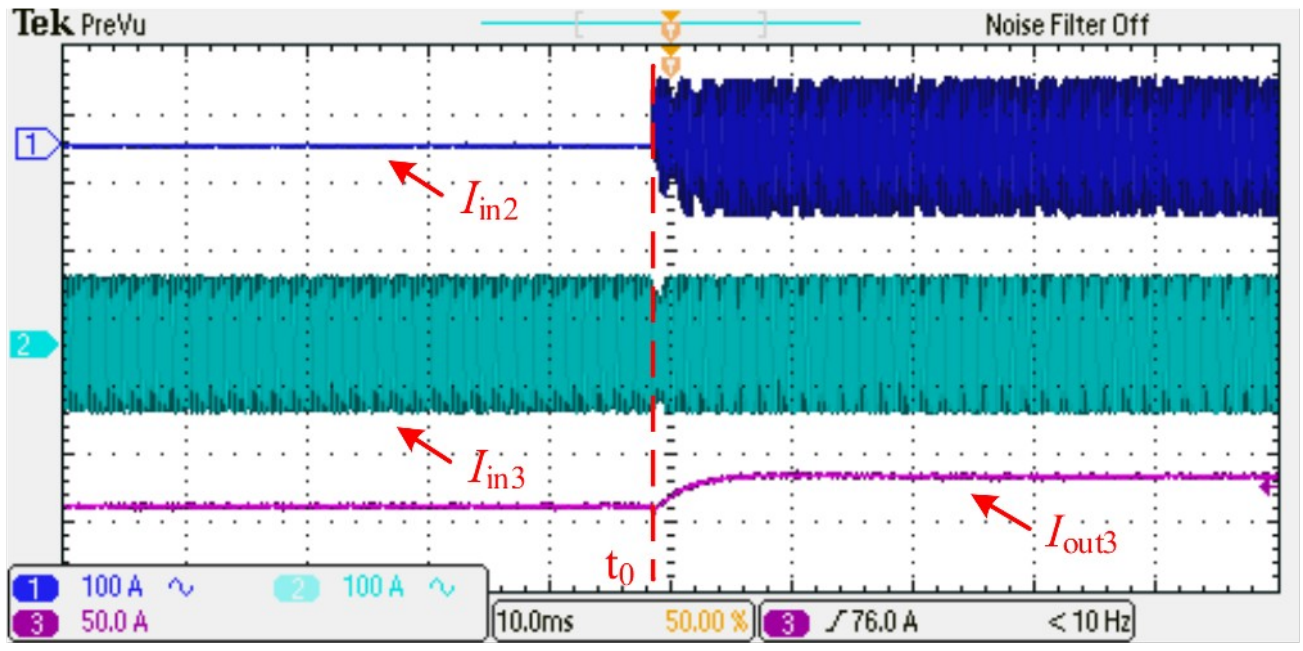

**Fig. 21.** Power sharing experiment

As shown in Fig. 21, the experimental waveform under power-sharing conditions demonstrates that when the battery is fully charged and the electrolyzer side requires more power, the auxiliary switch is activated at time $t_0$, allowing Ports 2 and 3 to simultaneously supply power to the electrolyzer. After time $t_0$, the input current at Port 2 recovers, and the current flowing into the electrolyzer unit significantly increases, proving the effectiveness of the auxiliary switch in achieving power sharing.

## VII. Conclusion

This paper proposes an open-circuit fault diagnosis and fault-tolerant scheme for a triple-port hydrogen production system and provides analysis and validation of the proposed approach.

1) The proposed fault diagnosis method utilizes the Sparrow Search Algorithm to optimize the Random Forest algorithm, thereby enhancing the speed and accuracy of fault diagnosis through the optimized Random Forest model.
2) The method requires only three current sensors, using the input current data $I_{in1}$, $I_{in2}$, and $I_{in3}$ from the three ports of the triple-port hydrogen production converter as fault diagnosis features.
3) The proposed fault diagnosis method can diagnose not only single-switch faults but also multiple-switch faults.
4) In the event of a switch fault, Implement fault-tolerant control within 0.05 ms-0.08 ms, the proposed scheme effectively ensures fault-tolerant operation while maintaining stable output power.
5) Power sharing enables flexible capacity scheduling between ports, increasing the power on the electrolyzer side, which further enhances the hydrogen production rate, making it more suitable for green hydrogen production.

This paper addresses the challenges faced in sustainable energy-based hydrogen production system by designing a triple-port hydrogen production converter with better economic efficiency and reliability, laying a solid foundation for the efficient and reliable production of green hydrogen.